\documentclass[conference]{IEEEtran}

\makeatletter
\newif\if@restonecol
\makeatother

\usepackage{subfigure}
\usepackage{url}
\usepackage{graphicx}
\usepackage{multirow}
\usepackage{booktabs}
\usepackage{array}
\usepackage{xspace}
\usepackage[linesnumbered,ruled,vlined]{algorithm2e}
\usepackage{algpseudocode}
\usepackage{amsmath}
\usepackage[colorlinks, linkcolor=blue, anchorcolor=blue, citecolor=blue]{hyperref}

\newcolumntype{L}[1]{>{\raggedright\let\newline\\\arraybackslash\hspace{0pt}}m{#1}}
\newcolumntype{C}[1]{>{\centering\let\newline\\\arraybackslash\hspace{0pt}}m{#1}}
\newcolumntype{R}[1]{>{\raggedleft\let\newline\\\arraybackslash\hspace{0pt}}m{#1}}

\begin{document}

\title{Cross-layer Path Selection in Multi-path Transport Protocol for Mobile Devices}

\author{\IEEEauthorblockN{Binbin Liao\IEEEauthorrefmark{1}\IEEEauthorrefmark{2}, Guangxing Zhang\IEEEauthorrefmark{1}, Qinghua Wu\IEEEauthorrefmark{1}, Zhenyu Li\IEEEauthorrefmark{1}\IEEEauthorrefmark{2} and Gaogang Xie\IEEEauthorrefmark{1}}
\IEEEauthorblockA{\IEEEauthorrefmark{1}ICT, CAS, China}
\IEEEauthorblockA{\IEEEauthorrefmark{2}University of CAS, China}
\{liaobinbin, guangxing, wuqinghua, xie\}@ict.ac.cn}

\maketitle


\begin{abstract}
MPTCP is a new transport protocol that enables mobile devices to use multiple physical paths simultaneously through several network interfaces, such as WiFi and Cellular. However, wireless path capacities change frequently in the mobile environments, causing challenges for path selection. For example, WiFi associated paths often become poor as devices walk away, since WiFi has intermittent connectivity caused by the short signal coverage and stochastic interference. MPTCP's native decision based on hysteretic TCP-layer estimation will miss the real switching point of wireless quality, which may cumulate packets on the broken path and causes serious packets reinjection. Through analyzing a unique dataset in the wild, we quantitatively study the impact of MAC-layer factors on the aggregated performance of MPTCP. We then propose a decision tree approach for cross-layer path selection that decides which path to carry the incoming packets dynamically according to the prior learned schemes. A prototype of the path selection system named SmartPS, which proactively probes the wireless environments, is realized and deployed in Linux and Android. Evaluation results demonstrate that our SmartPS can efficiently utilize the faster path, with goodput improvements of up to 29\%.
\end{abstract}
\section{Introduction}
\label{sec:intro}
The penetration of mobile devices such as smartphones and tablets has increased significantly over the last few years. This success is based on the rapid deployment and ubiquity of wireless technologies~\cite{DBLP:conf/pam/FukudaN13[5],chen2016cellular[6]}, such as IEEE 802.11 (WiFi) and cellular (4G/5G) communication networks. Many users with mobile devices can access the Internet through both WiFi and cellular networks. Typically, they only utilize one technology at a time: WiFi when it is available, and cellular otherwise. A multipath system such as MPTCP~\cite{ford2012tcp[1]} has focused on integrating various wireless access technologies to provide higher data rates, more services, and smooth handover modes~\cite{paasch2012exploring[2],du2015multipath[3]}. An MPTCP connection contains one or more subflows, each of which works the same as a regular TCP connection to the network. Starting with a four-way handshake, MPTCP's options are included in the SYN segments to verify whether the destination is MPTCP capable or not. And then a unique token number is negotiated to identify the first connection. Additional TCP subflows will be associated with this initial TCP subflows by carrying the previously token in their three-way handshake.

After designing an efficient path selector~\cite{paasch2013multipath[4]}, MPTCP can preferentially distribute the application data stream among the best subflows at the time to maximize the use of available physical interfaces. As only one TCP subflow will typically be associated to each physical interface, MPTCP's packet scheduler can improve the reaction to failures and maximize the connection-level throughput by injecting packets into the highest quality wireless interface. Thus, how to select which wireless interface to carry the incoming packets is becoming essential for the mobile device. MPTCP's native path decisions based on hysteretic TCP-layer estimation always miss the real switching point of wireless quality, which may cumulate packets on the broken path and causes serious packets reinjection.
\begin{figure}[htbp]
	\hspace*{0.15cm}
	\includegraphics[width=0.9\linewidth]{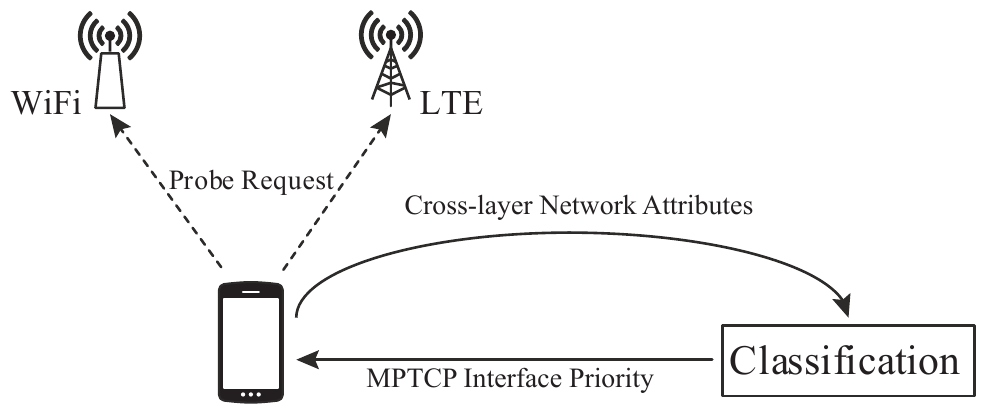}
	\caption{The process of path selection in SmartPS}
	\label{fig:process}
\end{figure}

Due to the symptom scaling property of TCP subflows, the QoS (Quality-of-Service) of the mobile application is critically subject to its wireless hop~\cite{mathis2008pathdiag[7]}. Especially when the mobile device walks away from a base station or access point, the short signal coverage with a fast signal fading has a significant impact on the wireless transmission~\cite{li2015measurement[8],li2018measurement[9]}. Several works (e.g.,~\cite{pefkianakis2015characterizing[10],patro2013observing[11]}) also show that serious interference in the available channels is a primary reason for the wireless quality decay in the wild. Naturally, various MAC-layer factors (e.g., RSSI, SINR) have proactively reflected these quality changes of the current wireless environment~\cite{mahajan2006analyzing[12],afroz2015sinr[14]}. However, there is no existing method integrates both TCP-layer factors and MAC-layer factors to make a cross-layer path selection.

In this paper, we carry out a quantitative analysis between cross-layer attributes and MPTCP connection-level performance, by inspecting a unique dataset collected from several mobile devices in the wild. We propose a wireless path selection approach by modeling it as a binary classification problem. We optimize the classification model by introducing random forest and post-pruning. These optimization improves classification accuracy, as well as reduces memory usage and computation cost. As shown in Fig.\ref{fig:process}, we implement SmartPS, a prototype of the path selection system based on Android with MPTCPv0.95. Evaluation results demonstrate that the proposed SmartPS can improve the goodput by up to 29 percent.

The rest of this paper is organized as follows. Section II discusses the background and motivation of MPTCP path selection. Section III introduces the dataset and analyzes the impact of various attributes. Section IV proposes a decision tree based approach with information gain ratio (IGR) and receiver operating characteristic (RoC). Section V describes the implementation of the sender side path selection system SmartPS and evaluates its performance in the real wireless environment. Section VI details related work and Section VII concludes our work.

\section{Background and Motivation}
\label{sec:background}
The rapid development of wireless networks has greatly changed the production of human life. A wide range of vertical use cases, such as massive Internet of things (IoT), remote machinery, autonomous driving, and virtual reality (VR), is expected by the next-generation mobile network~\cite{alliance20155g[22],deng2014ieee[23]}. However, limited by the open-air environment, the wireless network is more fickle and unstable than wired transmission. Especially in extreme situations~\cite{li2018measurement[9],cui2014fmtcp}, a mobile application relying on a single-path interface suffers great performance degradation. The recent multi-path schemes try to bridge the performance gap by integrating multiple wireless access technologies~\cite{paasch2012exploring[2],deng2014wifi[24],lim2017ecf[25]}. Because different wireless interfaces of a mobile device send or receive signals at different frequencies, which defines their signal coverage through the free-space path loss equation~\cite{elliot1983ieee[26]}. Moreover, the deployed base stations or access points may have different geographic elements, which also vary the background noise and multi-path effects of their wireless signal.

\begin{figure}[htbp]
\centering
\begin{minipage}[t]{0.45\linewidth}
\hspace*{-0.2cm}
\includegraphics[height=3cm,width=4.2cm]{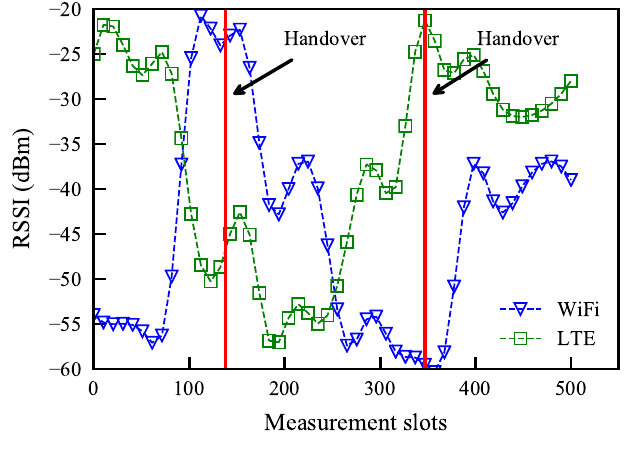}
\centerline{\footnotesize{Handover with the RSSI}}
\end{minipage}
\hspace*{0.3cm}
\begin{minipage}[t]{0.45\linewidth}
\hspace*{-0.3cm}
\includegraphics[height=3cm,width=4.2cm]{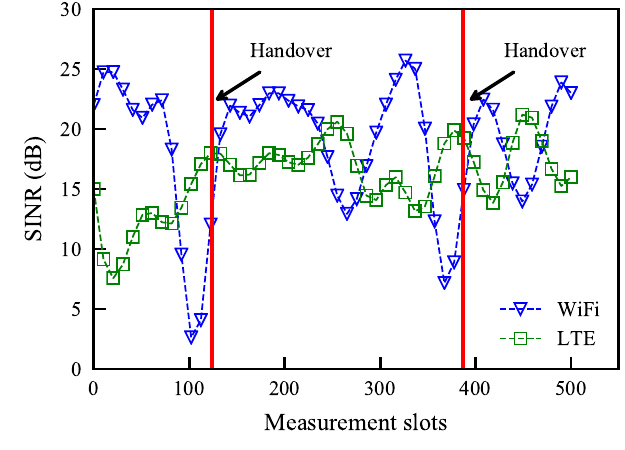}
\centerline{\footnotesize{Handover with the SINR}}
\end{minipage}
\caption{The relationship between MAC-layer attributes and handover point.}
\label{fig:fig2}
\end{figure}

\begin{figure}[htbp]
\centering
\begin{minipage}[t]{0.45\linewidth}
\hspace*{-0.2cm}
\includegraphics[height=3cm,width=4.2cm]{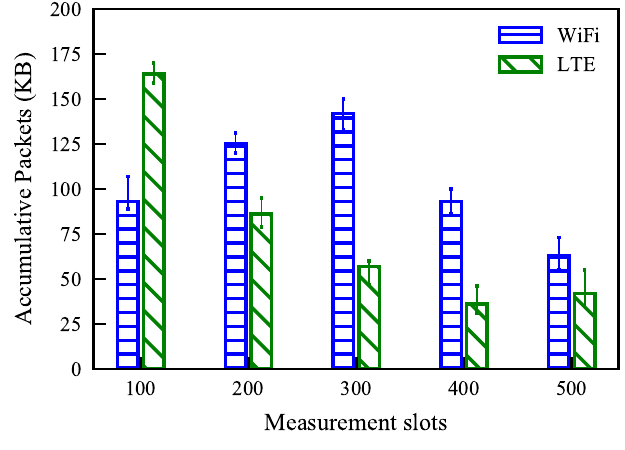}
\centerline{\footnotesize{\hspace*{0.2cm}Accumulation with the RSSI}}
\end{minipage}
\hspace*{0.3cm}
\begin{minipage}[t]{0.45\linewidth}
\hspace*{-0.3cm}
\includegraphics[height=3cm,width=4.2cm]{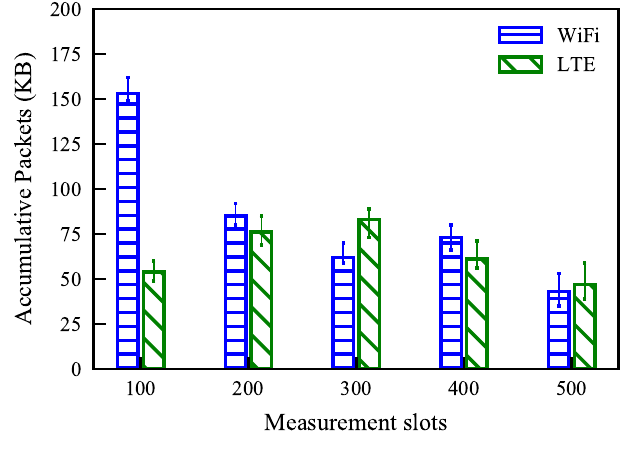}
\centerline{\footnotesize{\hspace*{0.2cm}Accumulation with the SINR}}
\end{minipage}
\caption{The accumulative in-flight packets at each measurement slots.}
\label{fig:fig3}
\end{figure}

Considering the low-speed mobility cases$\footnote{The speed is about 0$\sim$90 km/h}$, a good MPTCP path selector should proactively probe the wireless environment and dynamically return the best quality path to carry the incoming packets, as the mobile device may cross several network coverage regions$\footnote{Nanocell-5G or WiFi-5GHz has a shorter coverage}$ in a short time and result in frequent network quality switching. By selecting an appropriate path, the MPTCP sending buffer can significantly reduce the in-flight packets queue of the poor subflow, and improve the overall performance of the MPTCP connection. Through holding a general mobile device equipped with WiFi and LTE interfaces$\footnote{Dual LTE cards are common on Android device}$, we first explore the relationship between the representative MAC-layer attributes and the handover point of the default Minimum Round Trip Time First (MF) path selector. As the sampling result shown in Fig.\ref{fig:fig2}, we find that MF always misses the real switching point of wireless quality, since TCP-layer acknowledgement is delayed especially for a broken TCP subflow. And then we measure the in-flight packets of each subflow coupled with its physical interface. The result in Fig.\ref{fig:fig3} shows that the broken path accumulates more in-flight packets than the good one accompanying with the variation of MAC-layer attributes.

Several works indicate that leveraging the MAC-layer information of wireless interface can sense the quality of the network in advance~\cite{afroz2015sinr[14],sundaresan2015measuring[15]}. In order to enhance MPTCP's performance, Sinky et al. propose a proactive congestion control algorithm by detecting handoff with cross-layer assistance~\cite{sinky2016proactive[16]}. Lim et al. exploit MAC-Layer factors to estimate path status, and suspends/releases a path based on these features~\cite{lim2014cross[17]}. Palash et al. design a selective congestion retreat scheme with the received signal strength, which can suppress the weak TCP subflows without the need of explicit congestion notification~\cite{palash2018mpwifi[18]}. However, an arbitrary threshold is decided to manage the path in all this work, which includes such unreliable human experience that inapplicable to the general situation. By training a machine learning model, data-driven approaches with the information entropy have achieved great results in splitting decision factors~\cite{zhang2018data[19],pei2016wifi[20],li2018cutsplit[21]}. The above observation motivates us to design and implement a cross-layer selection of multiple wireless interfaces to improve MPTCP's aggregate performance based on a decision-tree method.
\begin{table*}[ht]
\centering
\caption{mac-layer attributes}
\label{Tab1}
\begin{tabular}{|c|c|c|c|l|}
\hline
\textbf{Abbreviation} & \textbf{MAC-layer Attributes}       & \textbf{Affiliation} & \textbf{Description}                              & \textbf{Unit} \\ \hline
RSSI                  & received signal strength indication & Both                 & the linear average of the total received power    & dBm           \\ \hline
SINR                  & signal to interference noise ratio  & Both                 & the ratio of the signal power to background noise & dB            \\ \hline
RSRP                  & reference signal received power     & LTE                  & the average power of the resource elements        & dBm           \\ \hline
RSRQ                  & reference signal received quality   & LTE                  & ranking among different candidate cells           & dB            \\ \hline
TD                    & transmitting data rate              & WiFi                 & physical bit rate for sending packets             & Mbps          \\ \hline
RD                    & receiving data rate                 & WiFi                 & physical bit rate for receiving packets           & Mbps          \\ \hline
\end{tabular}
\end{table*}

\section{Data Collection and Analysis}
\label{sec:data-analysis}
\subsection{MPTCP performance metrics}
Lots of assistant packet headers with different MPTCP options ensure the multipath transport procedures. However, only the encapsulated payloads are usable to the Application-layer. An ordinary throughput can not well measure the ability of MPTCP to aggregate traffic in the wireless network since packet loss and retransmission are common and cause numerous assistant packets. We thus use application goodput to indicate the delivery rate of valid packets, which neglects the extra overhead. As MPTCP ensures in-order delivery, the packets that are scheduled on the faster subflow have to wait for the slower subflow's packets to arrive in the out-of-order queue. We thus use application delay to indicate the packet delivery latency, which includes all the intermediate processing delays. According to the original intention of MPTCP design~\cite{raiciu2012hard[27]}, we define the better quality of an MPTCP connection as comparing its application goodput (AG) first and then comparing its application delay (AD) in the following section. The receiver-side host creates an iperf-session using the custom Aliyun cloud, which also compiles the latest MPTCP version to creates one subflow per physical interface for each MPTCP connection. The iperf-session runs for 60 seconds to allow the flows to reach equilibrium. We track the application delay by sending blocks of data, tagged with a timestamp~\cite{libpcap[13]}. Upon reception of each block, we store the transmission timestamp together with the timestamp at the receiver-side. The evolution of the difference between these timestamps gives us the variation in application delay.

\subsection{Cross-layer attributes}
Collection of the sensitive PHY layer parameters such as Channel State Information is hard and can only be obtained from few dedicated wireless chips, e.g., Intel 5300~\cite{halperin2011tool[28]}, Atheros 9580~\cite{xie2018precise[29]}. On the contrary, the general Android device provides convenient APIs~\cite{WifiInfo[30], telephony[31]} which enables us to monitor the basic parameters as the WiFi and Cellular MAC-layer attributes. Various information in Table~\ref{Tab1}, e.g., RSSI, SINR, RSRP, and RSRQ can comprehensively reflect the quality of the current LTE wireless environment. However, we must note that each physical interface is multiplexed by multiple protocols, although TCP traffic accounts for a large portion of the traffic. For each TCP subflow, its associated MAC-layer information are proactive but coarse-grained to affect MPTCP performance. Thus, discarding the single-factor dependence, we leverage more TCP-layer information, including RTT, CWND, Packet Delivery Rate (PDR), and Packet Loss Rate (PLR), to design selection judgments from various TCP subflows. Crossing the network layer, we also select the WiFi and Cellular MAC-layer information that have the largest correlation coefficients with the performance metrics and rewrite a $get\_mac\_opt()$ function with the above packages.

\subsection{Mobile device with multiple interfaces}
To analyze the status of cross-layer attributes, we collect a unique dataset from a real-world environment within a metropolis S, which has realized full coverage of emergent WiFi and 5G network in the urban area. We carried five different production Android devices built-in MPTCP, WiFi, and 4G/5G modules around more than 150 sampling locations, e.g., inside bars or restaurants. We have to point out that about a quarter of the time we are on the moving, either on foot or in the traffic. By pushing or pulling variable-sized web files, we measured the raw data exampled in Table~\ref{Tab2}, which lasts for 30 minutes in each place. The whole data collection lasts for 14 days, during which 1,246,742 valid samples in WiFi\&5G pair and 4,314,623 valid samples in WiFi\&4G pair are recorded. The $PRIO$ column determines whether the current path selection strategy is the WiFi path First (WF) or the LTE path First (LF). 
\begin{table}[!htbp]
\centering
\caption{MEDIAN VALUE OF KENDALL SCORE AND CONDITIONAL INFORMATION GAIN}
\label{Tab3}
\begin{tabular}{|c|c|c|c|c|}
\hline
\multirow{2}{*}{\textbf{Quality metric}} & \multicolumn{2}{c|}{\textbf{Kendall Score}} & \multicolumn{2}{c|}{\textbf{CIG}} \\ \cline{2-5} 
                                         & AG                   & AD                   & AG              & AD              \\ \hline
RSSI                                     & 0.61                 & 0.72                 & 0.12            & 0.23            \\ \hline
SINR                                     & 0.42                 & 0.39                 & 0.09            & 0.04            \\ \hline
RSRP                                     & 0.015                 & 0.017                 & 0.003           & 0.002           \\ \hline
RSRQ                                     & 0.013                & 0.022                & 0.004           & 0.006           \\ \hline
TD                                       & 0.005                & 0.001                & 0.007           & 0.003           \\ \hline
RD                                       & -0.003               & -0.001               & 0.003           & 0.005           \\ \hline
RTT                                      & 0.68                 & 0.73                 & 0.24            & 0.37            \\ \hline
CWND                                     & 0.34                 & 0.26                 & 0.12            & 0.08            \\ \hline
PLR                                      & -0.37                & -0.52                & 0.17            & 0.14            \\ \hline
PDR                                      & 0.78                 & 0.38                 & 0.31            & 0.11            \\ \hline
\end{tabular}
\end{table}

\begin{figure*}[!htbp]
\centering
\begin{minipage}[t]{0.3\linewidth}
\hspace*{-0.4cm}
\includegraphics[height=2.3cm,width=5.6cm]{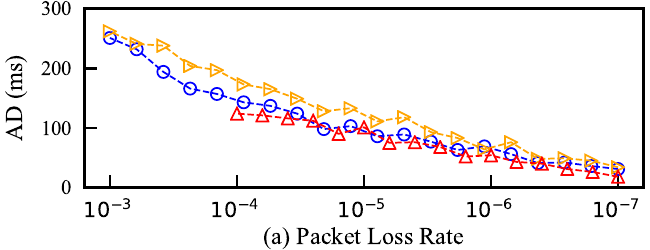}
\end{minipage}
\begin{minipage}[t]{0.3\linewidth}
\hspace*{-0.2cm}
\includegraphics[height=2.3cm,width=5.6cm]{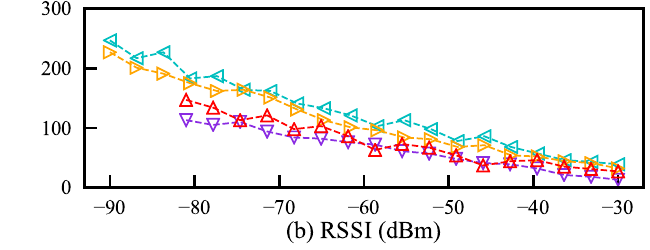}
\end{minipage}
\begin{minipage}[t]{0.3\linewidth}
\includegraphics[height=2.3cm,width=5.6cm]{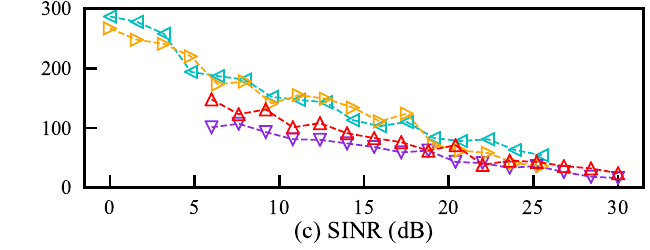}
\end{minipage}
\begin{minipage}[t]{0.3\linewidth}
\hspace*{-0.4cm}
\includegraphics[height=2.3cm,width=5.6cm]{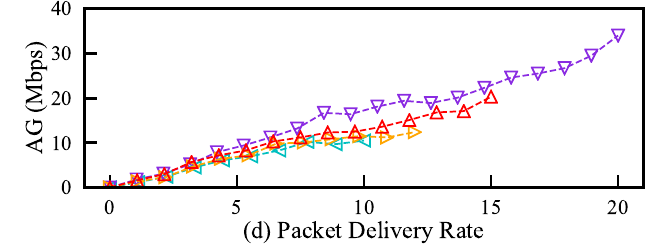}
\end{minipage}
\begin{minipage}[t]{0.3\linewidth}
\hspace*{-0.2cm}
\includegraphics[height=2.3cm,width=5.6cm]{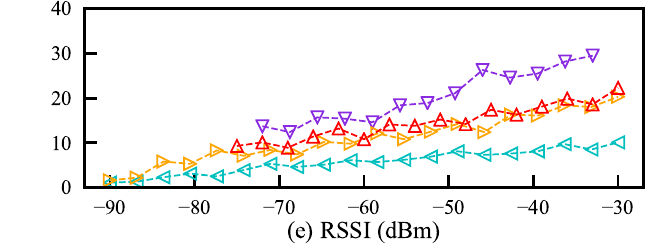}
\end{minipage}
\begin{minipage}[t]{0.3\linewidth}
\includegraphics[height=2.3cm,width=5.6cm]{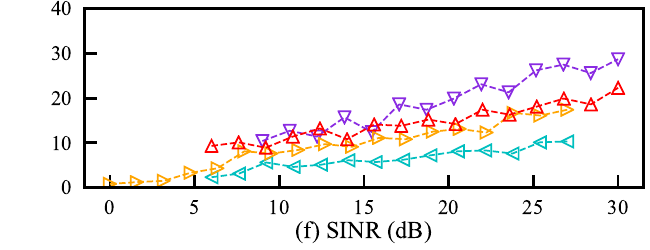}
\end{minipage}
\begin{minipage}[t]{0.3\linewidth}
\hspace*{-0.4cm}
\includegraphics[height=2.3cm,width=5.6cm]{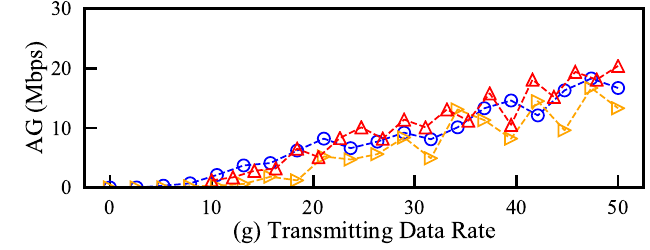}
\end{minipage}
\begin{minipage}[t]{0.3\linewidth}
\hspace*{-0.2cm}
\includegraphics[height=2.3cm,width=5.6cm]{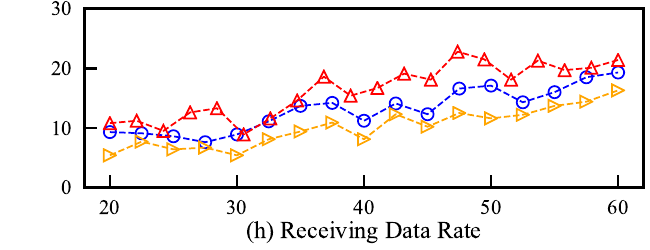}
\end{minipage}
\begin{minipage}[t]{0.3\linewidth}
\includegraphics[height=2.3cm,width=5.7cm]{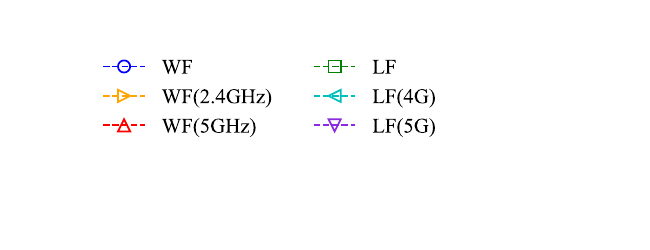}
\end{minipage}
\caption{Qualitative relationship between attributes and performance metrics on WF or LF.}
\label{fig4}
\end{figure*}

Next We plot Fig.~\ref{fig4} to understand how cross-layer factors influence the MPTCP performance metrics. Y-axis represents the corresponding performance metric, and X-axis in each subfigure represents a certain kind of attribute. we will not describe RTT and CWND in detail in this paper since they have been fully discussed in previous works~\cite{pei2016wifi[20],sinky2016proactive[16]}. We bin the attributes using different intervals based on their properties: 5dbm in RSSI, RSPR, 5dB in RSRQ, SINR, 5Mbps in TD, RD, PDR, and 0.05\% in PLR. The bins which contain less than 10 samples are filtered out to reduce the random errors. Then we use the 90th percentile statistical summary indicators for each bin's metrics. As the preferred interface will carry most of the MPTCP traffic, we divide the binning data into WF(5GHz), WF(2.4GHz), LF(5G), and LF(4G) based on their radio frequency difference. From Fig.~\ref{fig4}, we can find that MAC-layer attributes have a strong relationship with AG and AD. We further quantify the relationship by calculating the following two mathematical coefficients respectively.
\begin{table*}[!htbp]
\centering
\caption{raw dataset},
\label{Tab2}
\begin{tabular}{|c|c|c|c|c|c|c|c|c|c|c|c|c|c|c|c|}
\hline
\multicolumn{2}{|c|}{\textbf{RSSI}} & \multicolumn{2}{c|}{\textbf{SINR}} & \textbf{RSRP} & \textbf{TD} & \textbf{RSRQ} & \textbf{RD} & \multicolumn{2}{c|}{\textbf{PLR}} & \multicolumn{2}{c|}{\textbf{PDR}} & \multirow{2}{*}{\textbf{PRIO}} & \multirow{2}{*}{\textbf{AG}} & \multirow{2}{*}{\textbf{AD}} & \multirow{2}{*}{\textbf{LABEL}} \\ \cline{1-12}
LTE              & WiFi             & LTE             & WiFi             & LTE           & WiFi        & LTE           & WiFi        & LTE             & WiFi            & LTE             & WiFi            &                                &                              &                              &                                 \\ \hline
-39              & -29              & 17              & 22               & -62           & 3.7        & -12           & 13.2          & 0               & 0.03\%          & 8.3             & 4.5             & LF(4G)                         & 11.3                         & 47                           & \multirow{2}{*}{WF}             \\ \cline{1-15}
-42              & -27              & 12              & 19               & -58           & 9.4        & -14           & 22.5         & 0.01\%          & 0.04\%          & 3.4             & 16.2            & WF                             & 17.4                         & 35                           &                                 \\ \hline
-51              & -39              & 19              & 24               & -103          & 6.7        & -8            & 16.3         & 0               & 0.1\%           & 23.2            & 5.4             & LF(5G)                         & 24.1                         & 58                           & \multirow{2}{*}{LF}             \\ \cline{1-15}
-48              & -37              & 23              & 25               & -98           & 14.1        & -10           & 28.2        & 0               & 0.08\%          & 5.1             & 11.6            & WF                             & 13.4                         & 43                           &                                 \\ \hline
\end{tabular}
\end{table*}

To compare among different attributes, we first choose Kendall correlation as an indicator. This is because Kendall does not need to assume the data follows certain distribution (e.g., Gaussian distribution). Using the above bins of cross-layer attributes, we calculate the average value of corresponding application goodput, or delay that falls into this bin. Then we compute the Kendall correlation score between goodput or delay and different attributes for each physical interface pair. The median Kendall scores of different pairs are shown in the second column of Table~\ref{Tab3}. We can see that PDR, RTT, RSSI, and SINR have the largest correlation coefficient relative to AG. Meanwhile, RTT, RSSI, and PLR are the most relevant to AD.

However, correlation-based analysis cannot handle well those non-linear relationships such as RSSI, SINR, TD and RD in Fig. 4 (b), (c), (g), and (h). We choose conditional information gain, denoted as CIG, to resolve this problem. CIG represents how much of Y's uncertainty will be reduced with the knowledge of X's value. We first split the value of application goodput and delay into discretized values using 5Mbps and 5ms interval to format Y. For each kind of corresponding attribute Xi, we calculate the conditional information gain and repeat this for different interface pairs. The third column of Table~\ref{Tab3} shows PDR, RTT, RSSI have the top three CIG to AG. The definition of CIG, in this paper, represents the reduction percentage of AG or AD uncertainty after knowing the value of a certain cross-layer attribute, which ranges from 0 to 1. CIG equals 0 means that the performance metric is independent with a certain attribute. Any value of CIG which is larger than 0 shows that this attribute is useful for predicting performance metrics.

\subsection{Summary of data Analysis}
We can observe that both application goodput and application delay have a large Kendall or CIG with MPTCP's TCP-layer attributes (PDR, RTT). This indicates that the hysteretic network parameters can well reflect the current network conditions, which also means the wireless network we are connected is stable most of the time. This is also why the simple MF becomes the preferred strategy for the wired networks. However, we still need to consider the variability of wireless networks in mobile scenarios. Some MAC-layer attributes (RSSI,SINR) may have a larger Kendall and CIG than TCP-layer attributes (CWND,PLR), though only part of the data collection is carried out in an unstable environment. From the qualitative analysis, it is obvious that there is both a linear and a non-linear relationship between attributes and performance metrics. Moreover, the monotonic quantity of the linear relationship is quite different. It is almost impossible to accurately predict the performance Y by fitting a curve that covers all the attributes X. Fortunately, we only need to select WiFi or LTE based on the capturing attributes, which includes branch statements with partitioning problem. All we need to do is to know the threshold and the order in which the attributes are split.

\section{Modeling the Path Selection}
\label{sec:classification}
\subsection{Problem Statement}
For each pair of physical interfaces in Table~\ref{Tab2}, MPTCP's $PRIO$ choice for multiple paths causes a wide difference in the overall performance, although their attributes only have slightly different. If we take the $PRIO$ into account and group the data with the same priority before calculating the mathematical coefficient, e.g., only the attributes of WiFi are used to calculate Kendall and CIG when MPTCP selects the WiFi path first. After applying the same policy to the LTE path, the results in Table~\ref{Tab4} show a significant increase in Kendall score and CIG for most attributes. As the last column shown in Table~\ref{Tab2}, we merge the data rows of each pair by filling the median value of its two attributes and labeling the priorities with a better performance metric. After pre-processing the raw data, we get 5,564,365 entries in the classification dataset in total. The path selection problem is now modeled as a binary classification problem with Class LF and Class WF.
\begin{table}[!htbp]
\centering
\caption{PRIO-MEDIAN VALUE OF KENDALL SCORE AND CONDITIONAL INFORMATION GAIN}
\label{Tab4}
\begin{tabular}{|c|c|c|c|c|}
\hline
\multirow{2}{*}{\textbf{Quality metric}} & \multicolumn{2}{c|}{\textbf{Kendall Score}} & \multicolumn{2}{c|}{\textbf{CIG}} \\ \cline{2-5} 
                                         & AG                   & AD                   & AG              & AD              \\ \hline
RSSI                                     & 0.63                 & 0.74                 & 0.15            & 0.28            \\ \hline
SINR                                     & 0.44                 & 0.39                 & 0.11            & 0.06            \\ \hline
RSRP                                     & 0.02                 & 0.03                 & 0.004           & 0.002           \\ \hline
RSRQ                                     & 0.013                & 0.022                & 0.007           & 0.008           \\ \hline
TD                                       & 0.006                & 0.003                & 0.007           & 0.003           \\ \hline
RD                                       & -0.005               & -0.002               & 0.004           & 0.007           \\ \hline
RTT                                      & 0.73                 & 0.77                 & 0.26            & 0.43            \\ \hline
CWND                                     & 0.37                 & 0.27                 & 0.13            & 0.09            \\ \hline
PLR                                      & -0.43                & -0.54                & 0.18            & 0.15            \\ \hline
PDR                                      & 0.82                 & 0.39                 & 0.37            & 0.13            \\ \hline
\end{tabular}
\end{table}

\subsection{Modeling Approach}
Considering the complex relationship between cross-layer attributes and the performance metric, we choose machine learning (ML) as our tool to address the above problem. It is crucial to choose an appropriate ML method for accuracy. After a preliminary study, we choose Linear Regression, Naive Bayes, and Decision Tree (C5.0) from the Python scikit-learn package [18] as our classifier methods. Decision Tree method outperforms Linear Regression and Bayes methods in our pilot experiment. We infer that Bayes and Linear Regression have a poor performing on our dataset because the features in our dataset (i.e., MAC-layer attributes) are dependent on each other. The other reason is that the relationship between attributes and metrics is non-linear. However, the Decision Tree model has no assumptions about independence among features and linear relationship between features and labels.

\subsection{Evaluation}
We use 10-fold cross-validation to evaluate the accuracy of the Decision Tree models. We randomly divide the whole classification dataset into two parts: 90\% of data are used for training the models and the rest 10\% are used for evaluation. For the training procedures, we bin the samples by using the same method as Section~\ref{sec:data-analysis} and search the threshold partitions by comparing their Information Gain Ratio. The attributes RSRP, RSRQ, TD, and RD are excluded from the feature set because of their low CIG. We use these ground truth to build our Decision Tree with an iterative function. For the evaluation of ML models, we regard accuracy as the most important metric that reflects the overall performance of path selection. For the comparison, we add the classical ROC (Receiver Operating Characteristic) parameters, including precision, recall, and f1-score, to overemphasize the positive samples (WF in this paper). The results are summarized in Table~\ref{Tab5}. We do the classification on Linear Regression, Bayes, and Decision Tree respectively.
\begin{table}[htbp]
\centering
\caption{EVALUATION OF CLASSIFICATION ALGORITHMS}
\label{Tab5}
\begin{tabular}{|c|c|c|c|c|}
\hline
\multirow{2}{*}{\textbf{Classification Algorithm}} & \multicolumn{4}{c|}{\textbf{Evaluation Metrics}} \\ \cline{2-5} 
                                                   & Accuracy   & Precision   & Recall   & F1-score   \\ \hline
Linear Regression                                  & 0.427      & 0.441       & 0.397    & 0.418      \\ \hline
Naive Bayes                                        & 0.648      & 0.654       & 0.628    & 0.641      \\ \hline
Decision Tree (C5.0)                               & 0.828      & 0.831       & 0.814    & 0.822      \\ \hline
\end{tabular}
\end{table}

In this case, our Decision Tree achieves 82.8\% accuracy for classifying the preferential path, which performs better than Linear Regression and Bayes. The accuracy can be greatly improved by using the Random Forest model which votes between multiple trees. In our experiment, by using a Random Forest model whose tree number is 200, the accuracy can reach 86.2\%. The accuracy and Recall can further increase when increasing the tree number and the depth of the tree in the Random Forest model. However, it causes overfitting phenomenon and huge memory overhead because a large number of votable trees may generate a lot of unnecessary branches. To alleviate the problems, we can use a post-pruning method to merge the needless branch.
\begin{figure}[htbp]
\centering
\includegraphics[width=7.8cm,height=6.8cm]{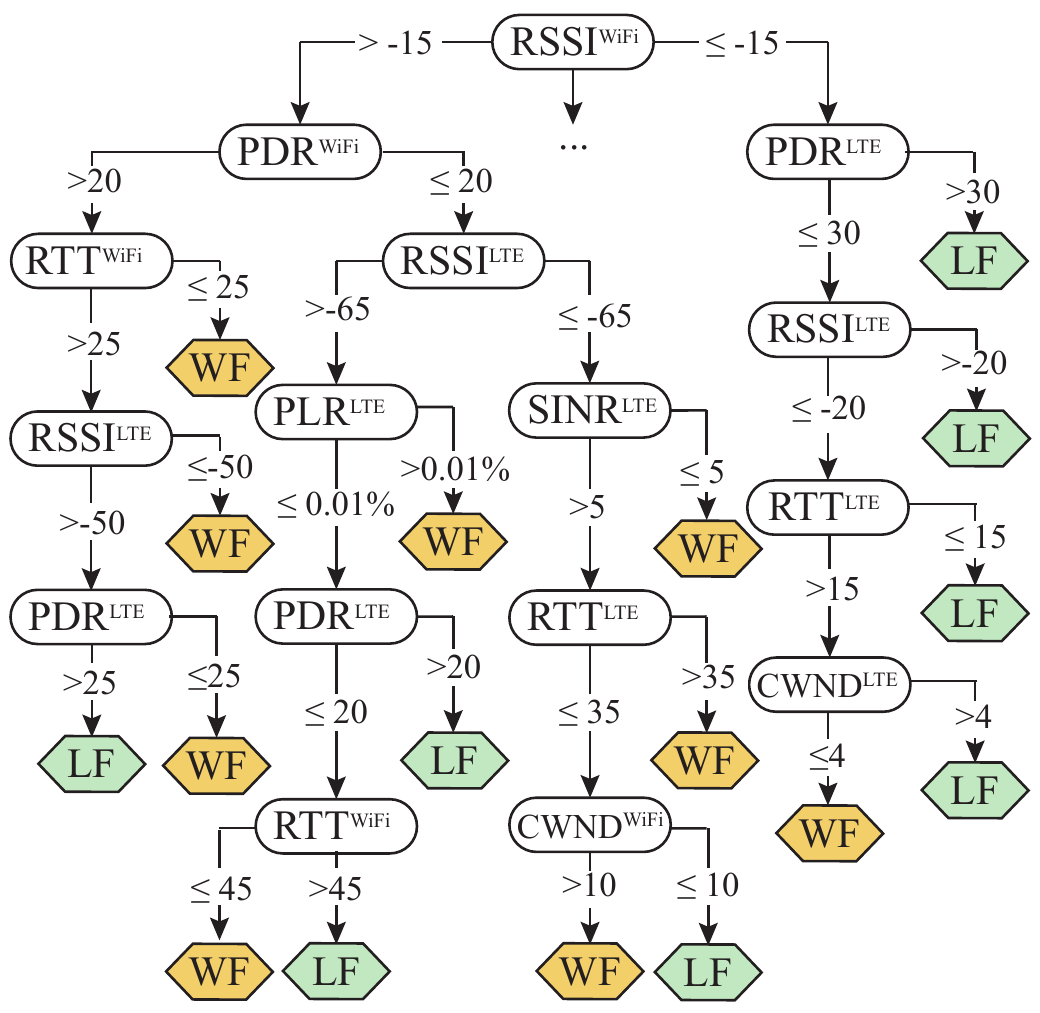}
\caption{Pruned Decision tree for classifying metrics into WF and LF.}
\label{fig5}
\end{figure}

As shown in Fig.~\ref{fig5}, we can know exactly what happened to the classified procedures by analyzing the partition nodes from root to a certain leaf. Because the learning model of Decision Tree in scikit-learn package is a local greedy algorithm, which confirms the splitting order and thresholds through comparing the entropy. RSSI is chosen as the root node of decision tree in Fig.~\ref{fig5} as $RSSI^{WiFi}$ has the largest information gain ratio during the startup. Meanwhile, the nodes in the remaining layers are determined by using the information gain ratio when combining their upper layer nodes. This combined information gain ratio used to purify the samples in each node differs from the overall information gain of single attribute in Table~\ref{Tab3}. This is why $SINR^{LTE}$ only have the 6th CIG but appear as the third layer of decision tree. In other words, the splitting order of $SINR^{LTE}$ right after $RSSI^{LTE}$ has a better branch gains than other explorations.

\section{Implementation and Evaluation of SmartPS}
\label{sec:prototype}
\subsection{Implementation of SmartPS}
In the previous section, we have trained a generic decision tree with cross-layer attributes, and optimized it by post-pruning heuristics and votable forest. To evaluate the performance of the modeled path selection in real wireless network, we must enforce our mobile device to select the multiple paths automatically based on the classification results. Considering the general situation, we denote that mobile devices only access the Internet through two kind of wireless network interfaces (i.e., WiFi and LTE). Even if other interfaces exist, we can actually retrain a multi-classification model by adding more attributes and labels. Although the server side has vast addresses available, we only consider the many-to-one case of mobile device mapping its peer server due to the presence of the load balancer, i.e., Each MPTCP connection establishes only one subflow on each physical interface. This not only simplifies our classification problem, but also avoids the performance degradation caused by the unfairness of the bottleneck~\cite{ferlin2016revisiting[32]}.
\begin{figure}[htbp]
\centering
\includegraphics[width=8.4cm, height=4.2cm]{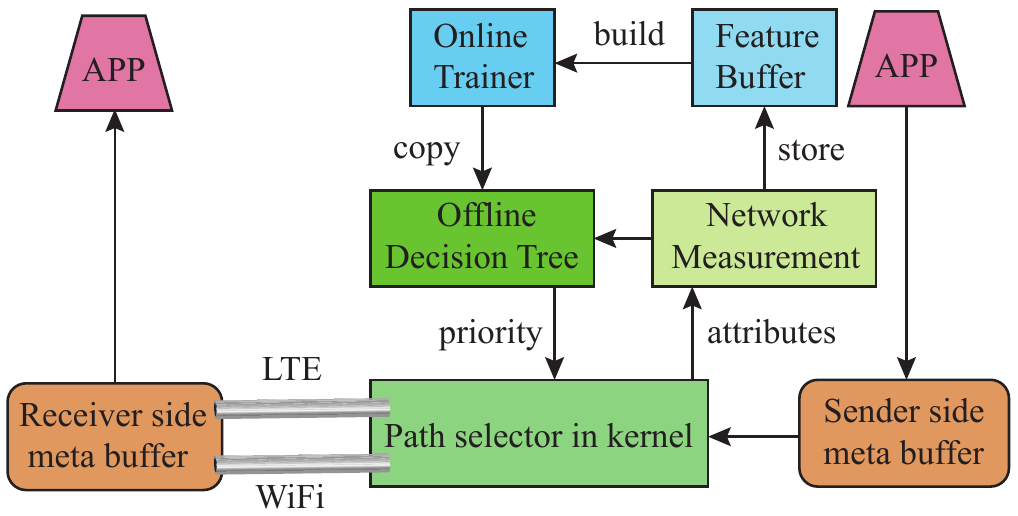}
\caption{System diagram of SmartPS}
\label{fig6}
\end{figure}

In the implementation of the SmartPS system, it requires interaction between the kernel and userspace as shown in Fig.\ref{fig6}. To simplify the execution logic, we rewrite and implement the path selector function~\cite{paasch2013multipath[4]} in the kernel that obtains the priority of each subflows from the decision tree in the userspace by a system call $setsockopt()$. Besides, we continuously leverage $getsockopt()$, $get\_mac\_opt()$, and iperf-session in the network measurement module to capture the cross-layer attributes and performance metrics. The detailed system settings are described as follows. As shown in Fig.~\ref{fig6}, to generate the realtime interface priority, SmartPS maintains two decision tree models. One well-trained offline decision tree is used to output the priority of the selected path according to the attributes of network measurement module. Another trainer builds an online decision tree according to the labeled dataset in the feature memory. To accommodate the changing wireless environment, the offline decision tree periodically copy its structure from the online module.

\subsection{Evaluation in commercial network}
To evaluate the actual effect of our decision tree-based path selection system, we choose the aforementioned five production Android devices that deployed in the wild to perform the experiment. By transferring files with variable size, we imitate the long/short MPTCP flows of different application services. Different means of transportation determine our speed of movement and the network environment along our trip. Using the sysctl command, we compared the SmartPS with MPTCP's native MinRTT and RoundRobin (RR) selector.
\begin{figure}[htbp]
\centering
\hspace*{-0.4cm}
\includegraphics[width=7.6cm,height=4cm]{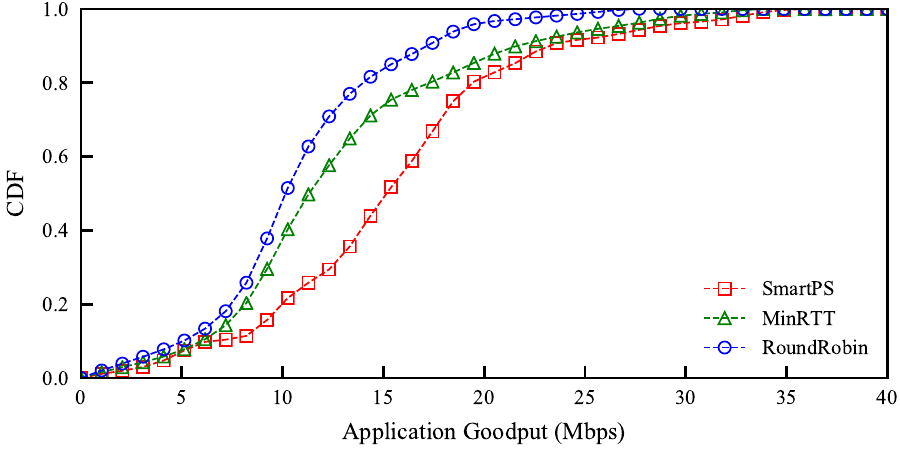}
\caption{The distribution of application goodput (AG).}
\label{fig6a}
\end{figure}
\begin{figure}[htbp]
\centering
\hspace*{-0.4cm}
\includegraphics[width=7.6cm,height=4cm]{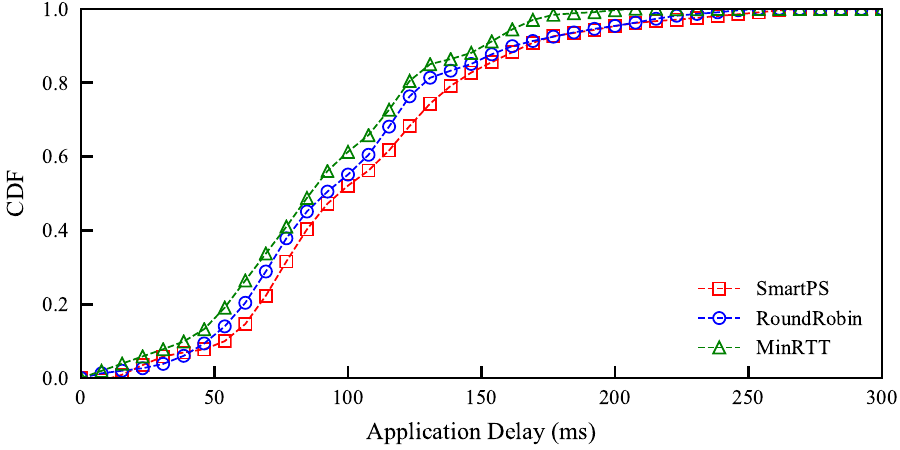}
\caption{The distribution of application delay (AD).}
\label{fig6b}
\end{figure}

\begin{figure}[htbp]
\centering
\hspace*{-0.4cm}
\includegraphics[width=6cm,height=4.32cm]{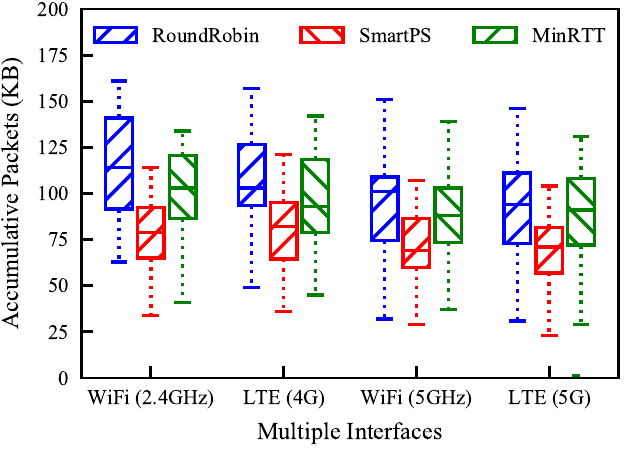}
\caption{The distribution of accumulative packets.}
\label{fig6c}
\end{figure}

Experimental results are shown in Fig.~\ref{fig6a} and Fig.~\ref{fig6b}. We first focus on the performance metrics, which indicate the quality of experience (QoE) of the  mobile application. Fig.~\ref{fig6a} shows the CDF of application goodput (AG) at the same mobile devices with different path selectors. SmartPS achieves a higher AG than MinRTT and RR in most of the time. The 50th percentile of AG in SmartPS is around 16.4 Mbps, but the 50th percentile of MPTCP connection in MinRTT and RR has AG no more than 12.7 Mbps, which makes about 29\% improvement. Because the interaction between Linux kernel and userspace, SmartPS may consume more time than the kernel-based path selector. Fig.~\ref{fig6b} presents the CDF of application delay (AD). As expected, SmartPS attaches 5.7\%-9.3\% AD than the simple MinRTT. We further measure the accumulative packets of multiple interfaces. SmartPS cumulates few bytes with more proaction in Fig.~\ref{fig6c}.

\subsection{Large-scale simulations}
Besides the real-world experiments, we also use Mininet to build a large-scale simulated testbed~\cite{paasch2013benefits[33]}. Based on the dataset in Section.~\ref{sec:data-analysis}, we create a virtual network and establish MPTCP connections between Mininet hosts. To test the extreme network environment, the domains of cross-layer attributes in each path not only cover the dataset but also randomly sets a wider range. SmartPS directly leverages these configurations without kernel interaction overhead.
\begin{figure}[htbp]
\centering
\hspace*{-0.4cm}
\includegraphics[width=7.6cm,height=4cm]{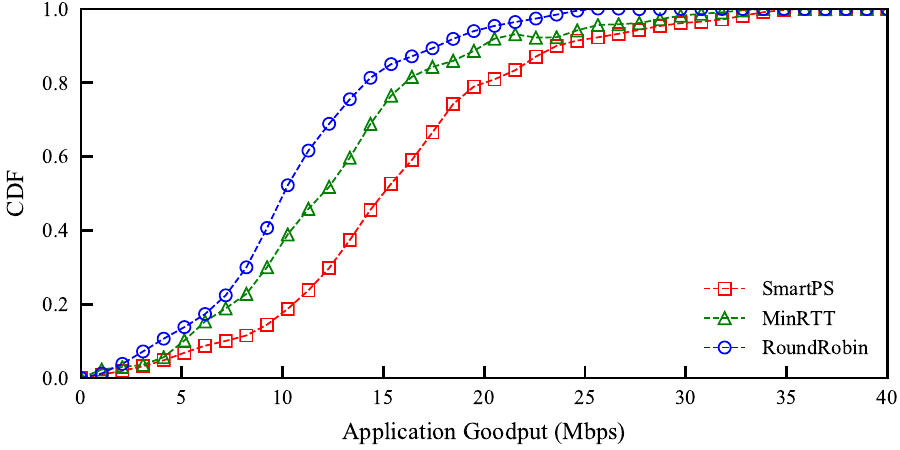}
\caption{The distribution of application goodput (AG).}
\label{fig7a}
\end{figure}
\begin{figure}[htbp]
\centering
\hspace*{-0.4cm}
\includegraphics[width=7.6cm,height=4cm]{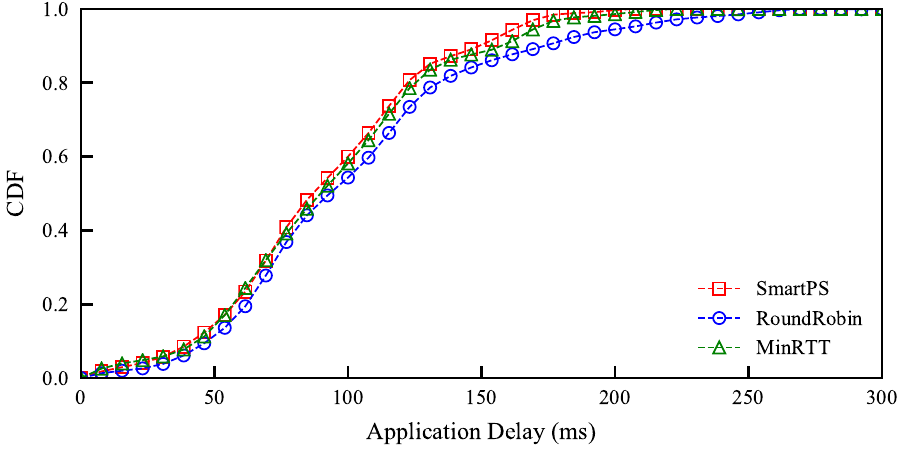}
\caption{The distribution of application delay (AD).}
\label{fig7b}
\end{figure}

\begin{figure}[htbp]
\centering
\hspace*{-0.4cm}
\includegraphics[width=6cm,height=4.3cm]{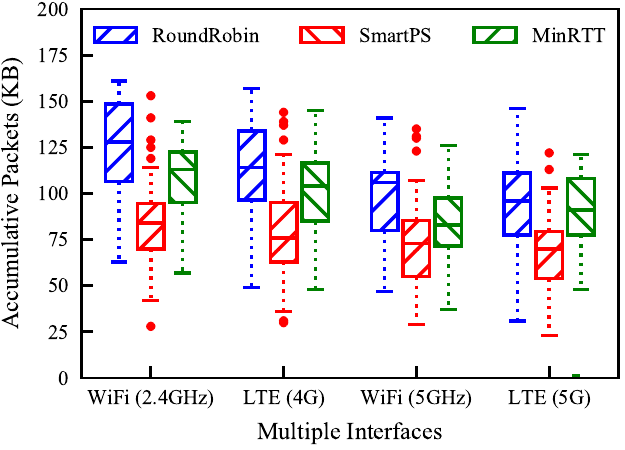}
\caption{The distribution of accumulative packets.}
\label{fig7c}
\end{figure}

As shown in the Fig.~\ref{fig7b}, our prototype reduces the 50th percentile of AD from 102 ms to 97 ms, about 4.8\%. However, it may suffer under-fitting problems and achieve a lower AG improvement than the real-world experiment, about 23.5\% in Fig~\ref{fig7a}. Fig.~\ref{fig7c} also shows this result that several outliers appear in the boxplots of accumulative packets.

\section{Related Work}
\label{sec:related}
The core idea of MPTCP path selection in our work is mostly inspired by~\cite{li2018cutsplit[21]}, which uses a general decision tree to support packet classification. Similarly, Zhang et al.~\cite{zhang2018data[19]} propose a decision tree approach for intelligent dual-band access selection. Zhou et al.~\cite{zhou2015demystifying[44]} structure the timeout retransmissions into a decision tree classification framework that identifies the cause of TCP performance issues. Leveraging the recent machine learning (ML) methods,~\cite{pei2016wifi[20],dobrian2011understanding[34]} and~\cite{balachandran2014modeling[35]} have detailed the relationship between WiFi/Cellular network attributes and mobile applications with a single-path TCP. Meanwhile, the popular Reinforcement Learning (RL)~\cite{sutton2018reinforcement[37]} combined with Long Short Term Memory (LSTM) neural network~\cite{hochreiter1997long[38]} has greatly improved the performance of MPTCP's scheduler~\cite{zhang2019reles[39]} and Congestion Controller~\cite{xu2019experience[40]}. Although MPTCP has evolved for many years, there is no previous work considering an efficient path selection based on cross-Layer information and appropriate ML.

Raiciu et al.~\cite{raiciu2011opportunistic[41]} have studied the mobility with MPTCP, which examined a mobile MPTCP architecture consisting of a mobile host, and an optional MPTCP proxy. While it shows MPTCP performs better stability in a mobile scenario, it does not examine the packet accumulative problem of broken subflow. Using different modes such as Full-MPTCP mode (where all potential subflows are used) and Backup mode (where only a subset of subflows are used to transmit packets), Passach et al.~\cite{paasch2012exploring[2]} discussed the impact of LTE/WiFi handover performance with MPTCP. However, they did not explore how to quickly utilize the faster path when path quality frequently changes.

Using the MAC-layer Information, Klemm et al.~\cite{klemm2005improving[43]} proposed mechanisms based on signal strength measurement to improve TCP performance in mobile ad-hoc networks. To alleviate packet losses due to mobility, their approach temporarily applies higher transmission power if the signal strength measurement indicates that a node is likely to move out of communication range. Li et al.~\cite{li2001link[42]} proposed Link Signal Strength Agent Protocol (LSSA) to report measured signal strength to the TCP-layer in mobile ad-hoc networks. However, it requires additional control packets to exchanges the path information with an unreliable single-path TCP connection. 

\section{Conclusion}
\label{sec:conclude}
In this paper, we presented the design, implementation, and evaluation of a decision tree-based framework, SmartPS, for path selector in MPTCP. We quantitatively analyze the cross-layer attributes that induce quality changes in LTE and WiFi, by inspecting a unique dataset collected from production Android devices in the wild. Using the attributes with a bigger information gain, we label the subflows' $PRIO$ for each MPTCP connection, by exploring and comparing the overall performance in two different priorities: WF or LF.

We propose a modification of MPTCP, called SmartPS, which selects subflows based on the standardized inputs capturing from cross-layer information. We leverage the random forest, online training, and post-pruning algorithms to optimize the trained decision tree which achieves higher application performance with lower memory usage and less computation cost. After implementing SmartPS in the Linux system on a cloud server and a mobile device, we perform experiments in the actual mobile scenarios and large-scale simulation. Our experimental results show that SmartPS is able to more quickly use the faster subflow than native MPTCP selector, and thus can achieve significantly better performance. For future work, we plan to extend the implementation of SmartPS into the userspace stack, which bypasses the Linux kernel and remove the bottleneck caused by kernel interaction.

\bibliographystyle{ieeetr}
\bibliography{reference}

\end{document}